\documentclass[conference]{IEEEtran}
\IEEEoverridecommandlockouts
\usepackage{cite}
\usepackage{amsmath,amssymb,amsfonts}
\usepackage{algorithmic}
\usepackage{graphicx}
\usepackage{textcomp}
\usepackage{xcolor}
\usepackage{subcaption}
\usepackage{hyperref}

\usepackage{booktabs}
\usepackage{makecell}

\usepackage{listings}
\usepackage{xcolor}
\def\BibTeX{{\rm B\kern-.05em{\sc i\kern-.025em b}\kern-.08em
    T\kern-.1667em\lower.7ex\hbox{E}\kern-.125emX}}

\begin{document}


\title{Towards Engineering Multi-Agent LLMs: A Protocol-Driven Approach}

\author{Zhenyu Mao$^{1}$, Jacky Keung$^{1}$, Fengji Zhang$^{1}$, Shuo Liu$^{1,*}$, Yifei Wang$^{1}$, and Jialong Li$^{2}$\\
	\normalsize $^{1}$City University of Hong Kong, Hong Kong, China $^{2}$Waseda University, Tokyo, Japan\\
	\normalsize zhenyumao2-c@my.cityu.edu.hk, Jacky.Keung@cityu.edu.hk, fzhang278-c@my.cityu.edu.hk \\
        \normalsize ywang4748-c@my.cityu.edu.hk, lijialong@fuji.waseda.jp\\
	\normalsize *corresponding author: sliu273-c@my.cityu.edu.hk 
}

\maketitle

\begin{abstract}

The increasing demand for software development has driven interest in automating software engineering (SE) tasks using Large Language Models (LLMs).
Recent efforts extend LLMs into multi-agent systems (MAS) that emulate collaborative development workflows, but these systems often fail due to three core deficiencies: under-specification, coordination misalignment, and inappropriate verification, arising from the absence of foundational SE structuring principles.
This paper introduces Software Engineering Multi-Agent Protocol (SEMAP), a protocol-layer methodology that instantiates three core SE design principles for multi-agent LLMs: (1) explicit behavioral contract modeling, (2) structured messaging, and (3) lifecycle-guided execution with verification, and is implemented atop Google’s Agent-to-Agent (A2A) infrastructure.
Empirical evaluation using the Multi-Agent System Failure Taxonomy (MAST) framework demonstrates that SEMAP effectively reduces failures across different SE tasks.
In code development, it achieves up to a 69.6\% reduction in total failures for function-level development and 56.7\% for deployment-level development.
For vulnerability detection, SEMAP reduces failure counts by up to 47.4\% on Python tasks and 28.2\% on C/C++ tasks.

\end{abstract}


\section{Introduction}
\label{sec:introduction}

As software has become an essential backbone supporting almost every aspect of modern society, nowadays there has been a rapid increase in demand not only for more software but also for more advanced software development to sustain and accelerate technological progress.
However, although the number of software developers is reaching 47 million in 2025 \cite{slashdata2025devs}, this remains insufficient to meet the rising demand, prompting increasing reliance on AI-powered tools.

Large Language Models (LLMs) have emerged as a promising solution to address the growing gap between software development demand and available developer resources.
With LLMs' strong ability to understand and interpret both natural and machine languages, LLMs have been successfully applied across a wide range of domains, including evolutionary computation \cite{cai2024exploring}, communication analysis \cite{fan2024coding}, self-adaptive systems \cite{li2024generative}, and, increasingly, software engineering (SE) \cite{SUN2025107743, 10.1145/3660778, mao2025hybrid}.
Recent research has explored how teams of LLMs can be orchestrated to simulate collaborative workflows in real-world SE tasks \cite{he2024llm} such as requirement engineering \cite{jin2024mare, ataei2024elicitronllmagentbasedsimulation}, code generation \cite{zhang2024pair, zan2024codesnaturallanguagecode}, quality assurance \cite{mao2024multi, hu2023largelanguagemodelpoweredsmart}, and maintenance \cite{wang2023intervenor, tao2024magisllmbasedmultiagentframework}.
In addition, several end-to-end multi-agent LLM frameworks, such as AutoGen, MetaGPT, and ChatDev, have been developed to model distinct development roles and workflows based on existing software process models such as Agile \cite{cohen2004introduction} and Waterfall \cite{petersen2009waterfall}.

Despite their conceptual appeal, current multi-agent LLM systems often under-perform in practice, as evidenced by high failure rates on SE tasks.
Cemri et al. \cite{cemri2025multi} introduced the Multi-Agent System Failure Taxonomy (MAST) and identified three recurring issue categories: under-specification, inter-agent misalignment, and inappropriate verification.

From a classical SE perspective, these failures can be interpreted as symptoms of three deeper structural deficiencies:
(1) inadequate component design, where agent responsibilities and role boundaries are poorly defined;
(2) insufficient interface specification, where inter-agent communication lacks semantic structure or typed formats; and
(3) inappropriate transition logic, where the system progresses between stages without formal gating or validation.
Traditional SE addresses such challenges through well-established principles: components are designed with explicit contracts, interfaces are formalized to ensure consistent integration, and system behavior is governed by state-based coordination models that enforce correctness through verification.
However, these SE principles remain largely absent from current multi-agent LLM frameworks, which are often built on loosely coupled prompts and informal, ad hoc role specifications.

To operationalize this vision, this paper introduces the Software Engineering Multi-Agent Protocol (SEMAP), a protocol-layer methodology that instantiates three core SE-inspired design principles: (1) explicit behavioral contract modeling, (2) structured messaging, and (3) lifecycle-guided execution with verification.
SEMAP is implemented as a lightweight protocol middleware atop Google’s Agent-to-Agent (A2A) infrastructure and supports both centralized and decentralized workflows.
While elements of these principles have appeared independently, SEMAP is the first to unify them in a domain-specialized protocol for the coordination in multi-agent LLMs.

\section{Background}
\label{sec:background}

\subsection{Multi-Agent LLMs for SE}

The recent progress in LLMs has led the development of multi-agent LLMs, where multiple LLM instances, each assuming a distinct, role-specific function, collaborate to solve complex tasks.
This approach is inspired by human collaborative work, in which modular responsibilities are distributed among team members and coordinated through structured communication.
In the context of SE, multi-agent LLMs seek to automate SE processes by breaking down high-level goals into sub-tasks and delegating them to specialized agents.

Multi-agent LLMs have been utilized to support a wide range of SE tasks.
In requirements engineering, they enable the elicitation, analysis, and validation of requirements by simulating human users interactions and processing natural language specifications \cite{ataei2024elicitronllmagentbasedsimulation, jin2024mare}.
For code generation, role-specialized agents, such as planners, implementers, and reviewers, collaborate to produce code, integrate components, and ensure alignment to coding standards \cite{zhang2024pair, zan2024codesnaturallanguagecode}.
Within the scope of quality assurance, multi-agent LLMs contribute to software testing through automated test case generation, execution, and analysis \cite{mao2024multi, hu2023largelanguagemodelpoweredsmart}.
Additionally, they support vulnerability detection by identifying and mitigating security flaws within software systems \cite{mao2024multi}.
In software maintenance, they assist in monitoring runtime behavior, diagnosing faults, and generating appropriate patches or modifications \cite{wang2023intervenor, tao2024magisllmbasedmultiagentframework}.


\subsection{Multi-Agent Protocols}

Recent efforts to formalize communication in multi-agent LLMs have led to the development of a diverse set of protocols.
A recent survey \cite{yang2025survey} categorized these protocols along two dimensions: (1) context-oriented vs. inter-agent, depending on whether the focus is on managing internal context of agents or external coordination between them, and (2) general-purpose vs. domain-specific, depending on whether they are intended for broad use or tailored to a specific domain.

A significant advancement in general-purpose agent interoperability is Google’s Agent-to-Agent (A2A) protocol.
A2A establishes a modular communication framework where autonomous agents interact through structured HTTP-based APIs, using JSON-RPC 2.0 as the foundational message format.
Each agent exposes a \textit{run()} function and publishes a machine-readable Agent Card, a JSON-based declaration of its identity, supported endpoints, input/output modalities (e.g., text, image, file), authentication requirements, and capabilities.
This approach enables dynamic discovery and seamless integration of agents across diverse environments, laying the groundwork for scalable and interoperable multi-agent LLMs.


\section{Methodology}
\label{sec:proposal}

\subsection{Methodology Overview}

\begin{figure*}[hbtp]
    \centering
    \includegraphics[width=150mm]{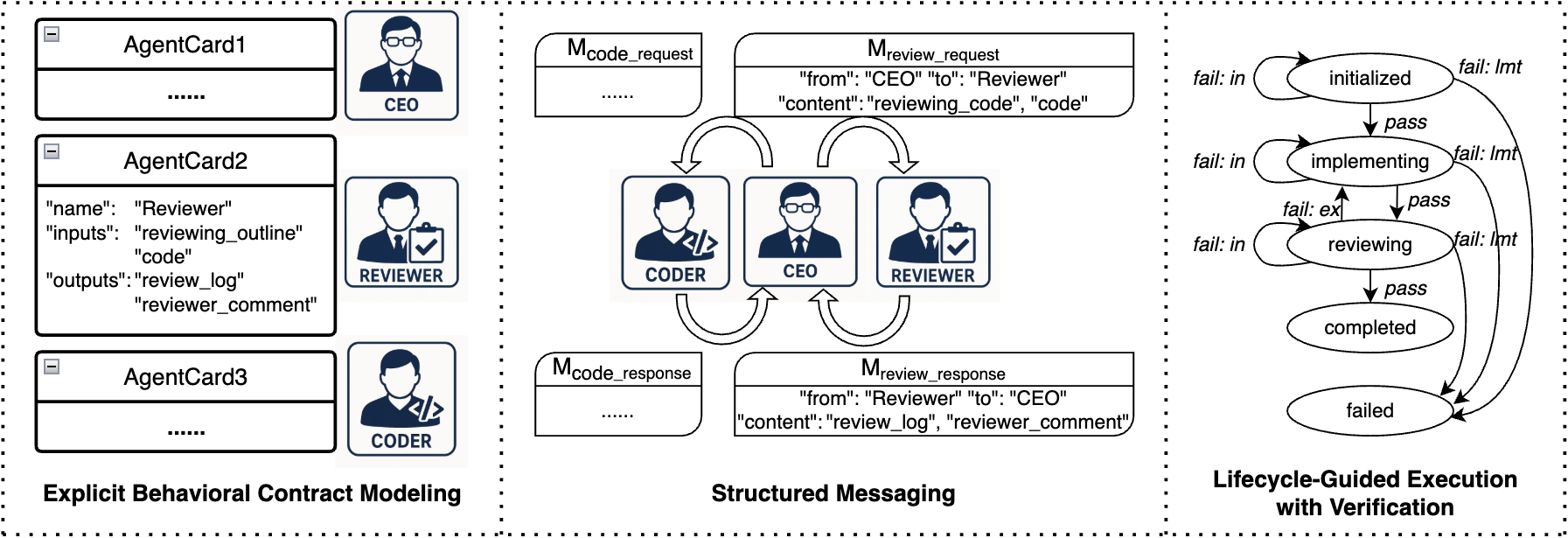}
    \caption{Methodology Overview}
    \label{fig:2}
\end{figure*}

This section introduces the Software Engineering Multi-Agent Protocol (SEMAP), a structured methodology built on the insight that reliable multi-agent coordination requires the same core abstractions that underpin classical SE, namely components, interfaces, and transition logic.
As shown in Fig.~\ref{fig:2}, SEMAP comprises three core principles, each grounded in foundational SE practices:
(1) \textit{Explicit behavioral contract modeling}, inspired by the principle of Design by Contract (DbC), formalizes agent responsibilities through pre-conditions and post-conditions. This reduces ambiguity and mitigates under-specification at both role and task levels;
(2) \textit{Structured messaging}, drawing from typed interface design and schema-driven communication in SE, enforces semantically typed inter-agent messaging to ensure clarity, completeness, and coordination alignment;
(3) \textit{Lifecycle-guided execution with verification}, reflecting state machine–based workflow modeling and stage-wise testing in SE, structures collaboration around a task-specific lifecycle with embedded verification gates. This ensures output correctness and guards against premature or invalid termination.

\subsection{Explicit Behavioral Contract Modeling}

To address under-specification in multi-agent LLMs, the proposed methodology adopts a contract-oriented modeling paradigm inspired by DbC principle.
In traditional SE, DbC establishes a formal agreement between software components, including pre-conditions define what must be true before execution, and post-conditions define what must hold afterward.
Translating this into the multi-agent LLM systems context, each agent is modeled through an explicit behavioral contract, a verifiable schema that specifies the required input artifacts and expected output artifacts.
Inputs capture the minimal artifacts the agent requires to operate meaningfully, such as task plans, peer feedback, or prior outputs, while outputs represent the agent’s expected contributions, such as the code artifact generated by the Coder, or a review log produced by the Reviewer.
This contract-oriented modeling formalizes agent responsibilities and role boundaries, reducing ambiguity across both design-time and runtime interactions.
Formally, a behavioral contract \( C \in \mathcal{C} \) is represented as a tuple as follows:
\[
C = \left(
\texttt{name},\ 
\mathcal{I}_C,\ 
\mathcal{O}_C
\right)
\]
where:
\begin{itemize}
    \item \( \texttt{name} \): a role identifier (e.g., Reviewer);
    \item \( \mathcal{I}_C \): set of required input artifacts (i.e., pre-conditions);
    \item \( \mathcal{O}_C \): set of required output artifacts (i.e., post-conditions).
\end{itemize}


\subsection{Structured Messaging}

While behavioral contracts define the roles, inputs and outputs of agents, effective coordination also requires that inter-agent communication be semantically clear and complete.
To address coordination misalignment in multi-agent LLMs, the proposed methodology adopts a structured messaging model that standardizes how information is exchanged between agents.
Each message \( M \in \mathcal{M} \) is formalized as:
\[
M = \left(
\texttt{sender},\
\texttt{receiver},\
\mathcal{C}_M
\right)
\]
where:
\begin{itemize}
    \item \( \texttt{sender} \): the identifier of the source agent;
    \item \( \texttt{receiver} \): the identifier of the target agent;
    \item \( \mathcal{C}_M \): a payload structured as a list of schema-designated objects (e.g., code, review\_log, reviewer\_comment).
\end{itemize}


\subsection{Lifecycle-Guided Execution with Verification}
\label{subsec:fsm}

While behavioral contracts and structured messaging ensure local correctness and coordination, they do not guarantee global task completion or output validity.
To address this, the methodology introduces a lifecycle-guided execution model, which structures agent collaboration as a state machine with verification-driven transitions.
This ensures that task progression is gated by validation and that failures can trigger appropriate recovery or reassignment actions.
Formally, a task lifecycle is modeled as a finite state machine (FSM):
\[
\mathcal{L} = (\mathcal{S}, \Sigma, \delta, s_0, \mathcal{F})
\]
where:
\begin{itemize}
    \item \( \mathcal{S} \): a set of lifecycle stages (e.g., initialized, implementing, reviewing, completed, failed);
    \item \( \Sigma \): verification outcomes (e.g., pass, fail);
    \item \( \delta: \mathcal{S} \times \Sigma \rightarrow \mathcal{S} \): a transition function for the next stage;
    \item \( s_0 \in \mathcal{S} \): the initial stage (typically initialized);
    \item \( \mathcal{F} \subseteq \mathcal{S} \): terminal stages (e.g., completed, failed).
\end{itemize}

\section{Preliminary Evaluation}
\label{sec:evaluation}

This section presents the evaluation of SEMAP, focusing on its effectiveness in reducing failures in multi-agent LLMs for SE tasks.
The research questions (RQs) are set as follows:
\begin{itemize}
    \item RQ1: To what extent does SEMAP mitigate failures across the three major categories, namely under-specification, coordination misalignment, and verification failure, compared to the baseline in different SE tasks?
    \item RQ2: How effectively does SEMAP enable consistent and stable failure reduction across collaboration rounds, particularly in minimizing recurrence in each category?
\end{itemize}

\subsection{Experiment Settings}

\paragraph{Tasks and datasets}

The experiments include two representative SE tasks: software development and vulnerability detection, each evaluated using two comprehensive datasets.

\textbf{Function-level development:}
In function-level development, agents solve problems from the HumanEval \cite{chen2021evaluating}, each requiring the writing of a concise function that satisfies a given textual specification.

\textbf{Deployment-level development:}
The deployment-level development task is based on ProgramDev \cite{cemri2025multi}, where agents develop a complete deployment starting from a one-sentence user requirement.

\textbf{C/C++ vulnerability detection:}
This task uses \texttt{devign100}, a 100-sample subset of the Devign dataset \cite{zhou2019devign}, constructed by randomly selecting 50 vulnerable and 50 safe C/C++ functions.
Each sample contains a function-level code snippet and a binary label indicating the presence or absence of a vulnerability.
The code snippets are typically short (under 30 lines), enabling agents to focus on localized vulnerability patterns such as pointer misuse or unsafe memory operations.

\textbf{Python vulnerability detection:}
This task uses \texttt{vudenc100}, a 100-sample dataset created by randomly sampling from the CVEFixes dataset \cite{TRAN2024107504}.
A function is labeled as vulnerable if any of its lines are marked as such in the original line-level annotations.
Similarly, the final dataset consists of 50 vulnerable and 50 safe Python functions.
Compared to \texttt{devign100}, these code snippets are longer and structurally more complex, often containing real-world CVE patches with nested logic and broader reasoning scopes.

\paragraph{Baseline and model settings}

The baseline system is implemented using the MetaGPT framework.
For development tasks, it adopts a centralized CEO-style multi-agent architecture consisting of five agents: a CEO, a Planner, a Coder, a Reviewer, and a Tester.
For vulnerability detection tasks, both the baseline and SEMAP use a decentralized three-agent settings consisting of an Auditor, a Critic, and a Tester.

Experiments are conducted using DeepSeek-V3-0324 and gpt-4.1-nano-2025-04-14 as the underlying agent models.
For development tasks, SEMAP and the baseline are allowed up to five collaboration rounds per task. For vulnerability detection, the system executes a single round of analysis and voting.

\paragraph{Evaluation metrics}

The LLM-as-a-Judge pipeline proposed in \cite{cemri2025multi} is employed to categorize multi-agent LLMs failures, assigning each instance to one or more categories: specification issues, inter-agent misalignment, and task verification failures, using a separate model, gpt-4o-2024-08-06.


To evaluate RQ1, total number of failures in each category across different SE tasks, including both development and vulnerability detection, is compared.
To evaluate RQ2, change in failure counts change across collaboration rounds in development tasks, is compared between SEMAP and baseline.

\subsection{Experiment Results}

\textbf{Results for RQ1:} The total number of failures in each of the three major categories are reported in Table \ref{table:2} (development tasks) and Table \ref{table:3} (vulnerability detection), respectively.
Across all datasets and model configurations, SEMAP consistently reduces the number of failures compared to the baseline.

In function-level development (HumanEval), SEMAP reduces the total number of failures by 64.1\% with ChatGPT (from 256 to 92) and by 69.6\% with DeepSeek (from 112 to 34).
The largest reduction occurs in under-specification, where ChatGPT drops from 137 to 39 (71.5\%), and DeepSeek from 63 to 17 (73.0\%). Similarly, in deployment-level development (ProgramDev), SEMAP achieves a 12.6\% reduction with ChatGPT (from 103 to 90) and a 56.7\% reduction with DeepSeek (from 67 to 29), with the most notable improvement in under-specification for DeepSeek (39 to 18, a 53.8\% decrease) and a complete elimination of inter-agent misalignment errors.

In vulnerability detection, SEMAP also shows consistent advantages over the baseline.
On \texttt{devign100}, total failures decrease by 28.2\% with ChatGPT (from 78 to 56) and by 8.3\% with DeepSeek (from 48 to 44). 
On \texttt{vudenc100}, the reduction is 47.4\% with ChatGPT (from 38 to 20) and 16.4\% with DeepSeek (from 55 to 46).
Unlike development tasks, the reductions here are more evenly distributed across the three failure categories, indicating SEMAP’s balanced impact on decentralized workflows involving parallel agent collaboration.

\textit{Thus, RQ1 is answered: SEMAP consistently reduces failure counts across all categories and tasks, with the most significant improvements observed in under-specification during development, and more balanced reductions in vulnerability detection.}

\begin{table*}[htbp]
\centering
    \caption{Results of RQ1: Development Tasks} 
    \begin{tabular}{ccccccccccccc}\hline 
         & \multicolumn{6}{c}{HumanEval} & \multicolumn{6}{c}{ProgramDev}\\
        Failure Category & \multicolumn{3}{c}{GPT-4.1-nano} & \multicolumn{3}{c}{Deepseek-V3} & \multicolumn{3}{c}{GPT-4.1-nano} & \multicolumn{3}{c}{Deepseek-V3}\\
         & Baseline & SEMAP & $\Delta\%$ & Baseline & SEMAP & $\Delta\%$ & Baseline & SEMAP & $\Delta\%$ & Baseline & SEMAP & $\Delta\%$\\\hline
        Under-specification & 137 & 39 & 71.5 & 63 & \textbf{17} & 73.0 & 48 & 46 & 4.1 & 39 & \textbf{18} & 53.8 \\
        Inter-Agent Misalignment & 27 & 5 & 81.5 & 10 & \textbf{3} & 70.0 & 9 & 9 & 0.0 & 8 & \textbf{0} & 100.0\\
        Task Verification & 92 & 48 & 47.8 & 39 & \textbf{14} & 64.1 & 46 & 35 & 23.9 & 20 & \textbf{11} & 45.0 \\\hline
        Total & 256 & 92 & 64.1 & 112 & \textbf{34} & 69.6 & 103 & 90 & 12.6 & 67 & \textbf{29} & 56.7 \\\hline
    \end{tabular}
    \label{table:2}
\end{table*}

\begin{table*}[htbp]
\centering
    \caption{Results of RQ1: Vulnerability Detection Tasks} 
    \begin{tabular}{ccccccccccccc}\hline 
         & \multicolumn{6}{c}{devign100 (C/C++)} & \multicolumn{6}{c}{vudenc100 (Python)}\\
        Failure Category & \multicolumn{3}{c}{GPT-4.1-nano} & \multicolumn{3}{c}{Deepseek-V3} & \multicolumn{3}{c}{GPT-4.1-nano} & \multicolumn{3}{c}{Deepseek-V3}\\
         & Baseline & SEMAP & $\Delta\%$ & Baseline & SEMAP & $\Delta\%$ & Baseline & SEMAP & $\Delta\%$ & Baseline & SEMAP & $\Delta\%$\\\hline
        Under-specification & 12 & 9 & 25.0 & 5 & \textbf{4} & 20.0 & 4 & \textbf{3} & 25.0 & 16 & 15 & 6.3 \\
        Inter-Agent Misalignment & 33 & 24 & 27.2 & 20 & \textbf{18} & 10.0 & 14 & \textbf{11} & 21.3 & 16 & 11 & 31.3\\
        Task Verification & 33 & 23 & 30.3 & 23 & \textbf{20} & 13.0 & 10 & \textbf{6} & 40.0 & 23 & 20 & 13.0 \\\hline
        Total & 78 & 56 & 28.2 & 48 & \textbf{44} & 8.3 & 38 & \textbf{20} & 47.4 & 55 & 46 & 16.4 \\\hline
    \end{tabular}
    \label{table:3}
\end{table*}

\begin{figure*}[htbp]
    \centering
    
    \begin{subfigure}[t]{0.32\textwidth}
        \includegraphics[width=\linewidth]{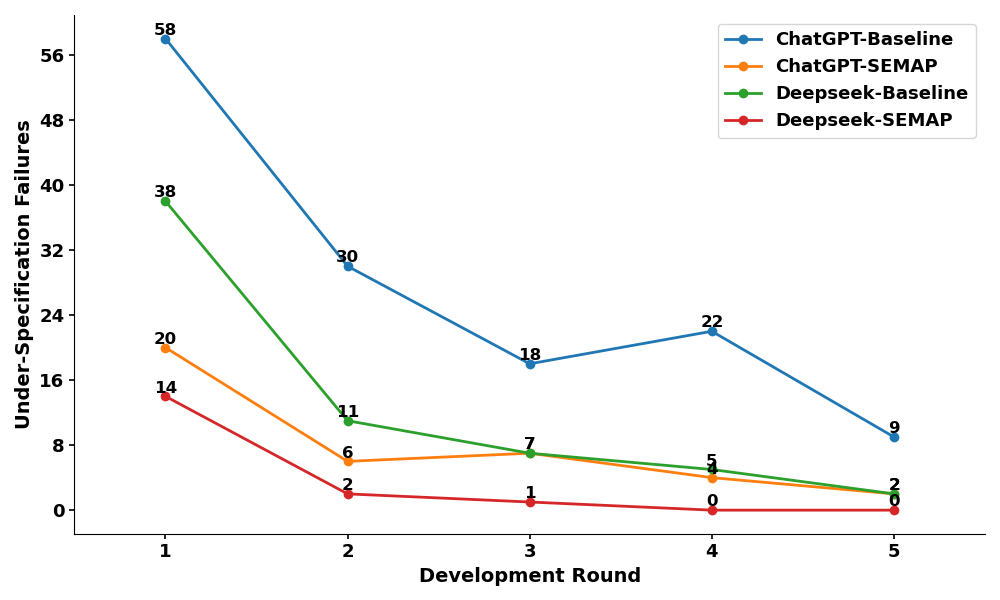}
        \caption{HumanEval: Under-Specification}
        \label{fig:3a}
    \end{subfigure}
    \begin{subfigure}[t]{0.32\textwidth}
        \includegraphics[width=\linewidth]{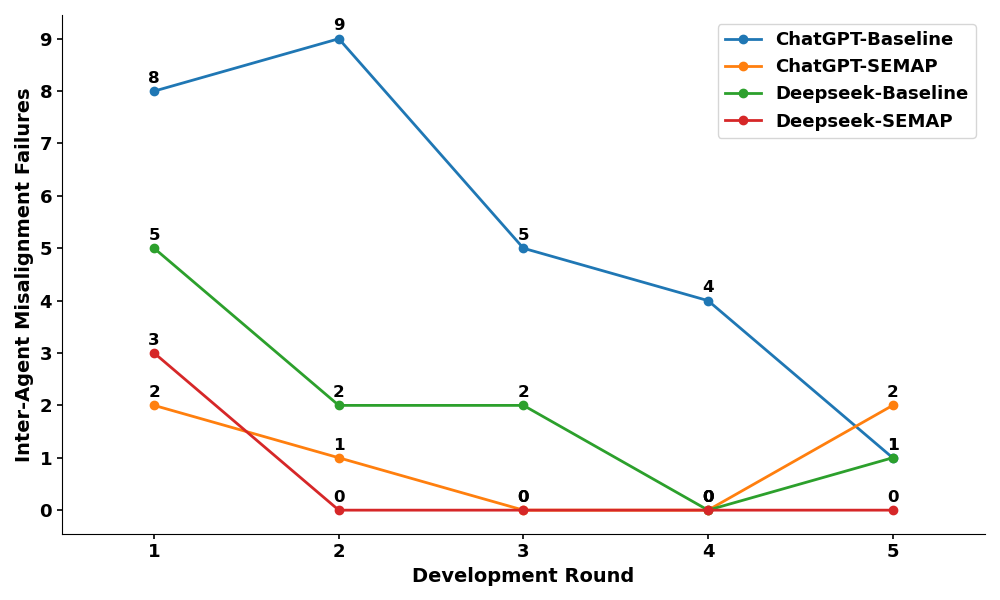}
        \caption{HumanEval: Inter-Agent Misalignment}
        \label{fig:3b}
    \end{subfigure}
    \begin{subfigure}[t]{0.32\textwidth}
        \includegraphics[width=\linewidth]{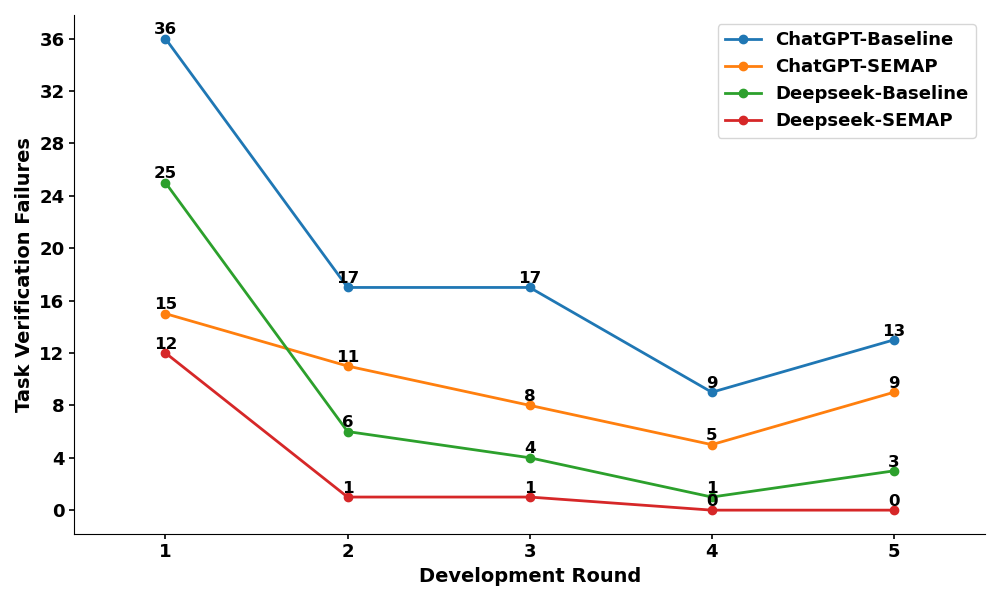}
        \caption{HumanEval: Task Verification}
        \label{fig:3c}
    \end{subfigure}
    
    \begin{subfigure}[t]{0.32\textwidth}
        \includegraphics[width=\linewidth]{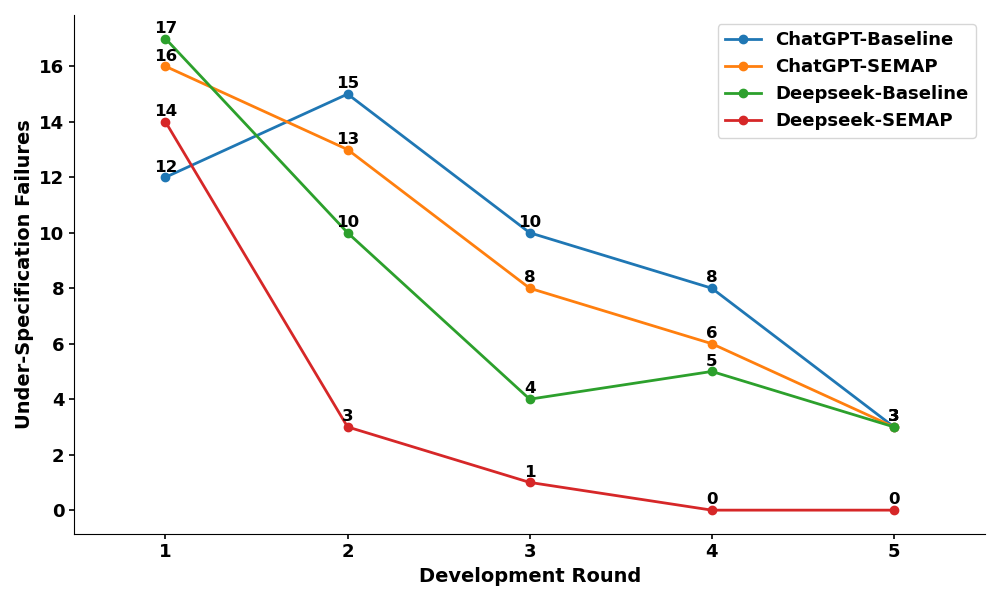}
        \caption{ProgramDev: Under-Specification}
        \label{fig:3d}
    \end{subfigure}
    \begin{subfigure}[t]{0.32\textwidth}
        \includegraphics[width=\linewidth]{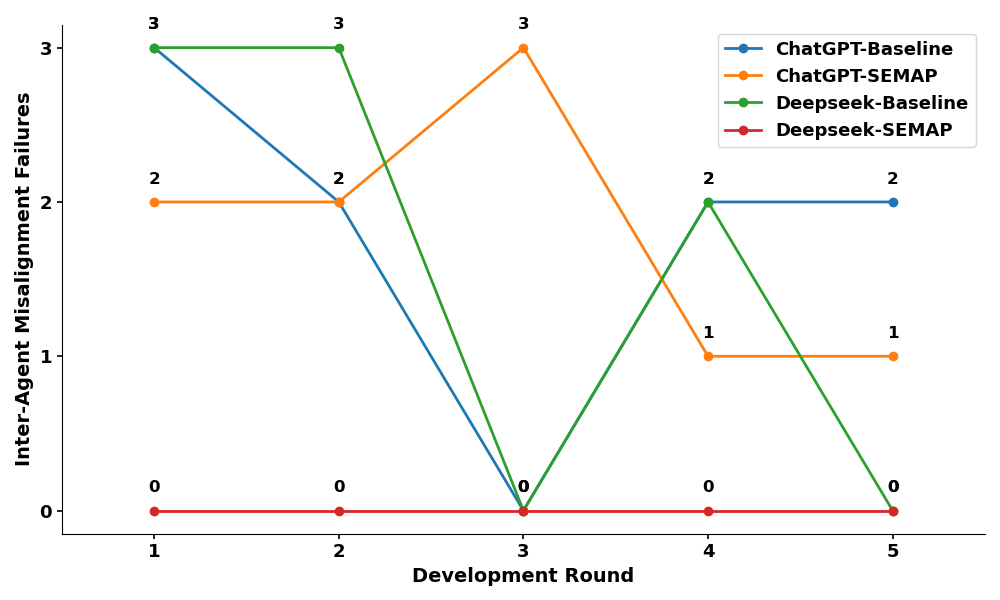}
        \caption{ProgramDev: Inter-Agent Misalignment}
        \label{fig:3e}
    \end{subfigure}
    \begin{subfigure}[t]{0.32\textwidth}
        \includegraphics[width=\linewidth]{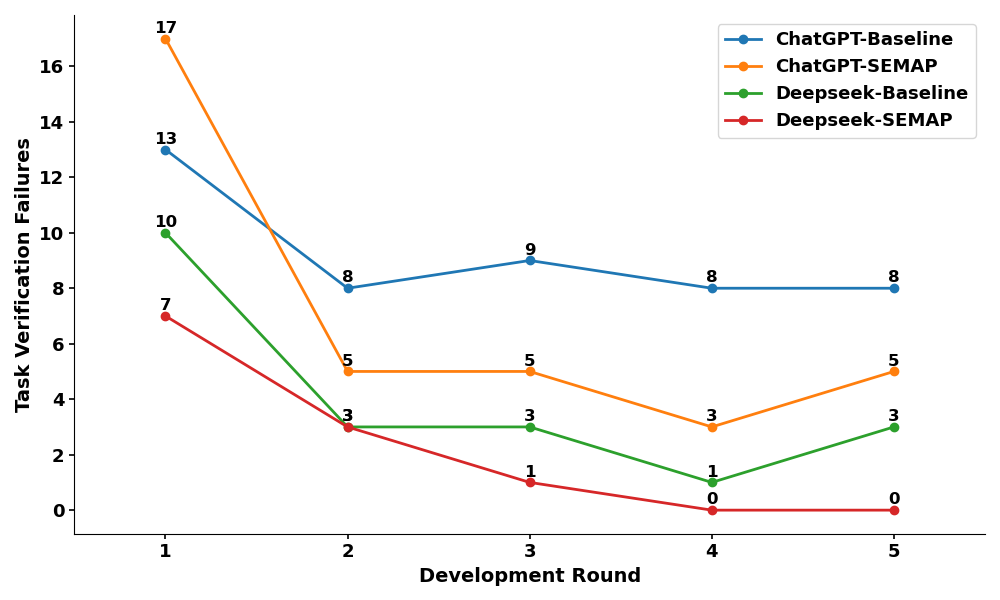}
        \caption{ProgramDev: Task Verification}
        \label{fig:3f}
    \end{subfigure}
    
    \caption{Results of RQ2: Failures v.s. Rounds in Development Tasks}
    \label{fig:3}
\end{figure*}

\textbf{Results for RQ2.} Figure \ref{fig:3} illustrates how failure counts evolve across development rounds, separated by different datasets and failure categories.
Unlike RQ1, which evaluates end-to-end failure counts, RQ2 focuses on its changes during the iterative process, examining whether SEMAP promotes more stable and robust collaboration over repeated trials.

In the HumanEval dataset, SEMAP results in fewer failures across all development rounds compared to the baseline.
As shown in Fig. \ref{fig:3a}, SEMAP leads to a sharp and steady reduction in under-specification failures, dropping from 20 to 2 for ChatGPT-SEMAP and from 14 to 0 for DeepSeek-SEMAP by round 5.
However, the baseline declines more gradually, with ChatGPT-Baseline ending at 9 failures and DeepSeek-Baseline at 2.
Similar patterns also can be found in inter-agent misalignment (Fig. \ref{fig:3b}) and task verification (Fig. \ref{fig:3c}).

In the ProgramDev dataset, SEMAP continues to demonstrate improved round-wise behavior compared to the baseline.
As shown in Fig. \ref{fig:3d}, SEMAP significantly reduces under-specification failures across rounds, with DeepSeek-SEMAP dropping from 14 to 0 and ChatGPT-SEMAP from 16 to 3 by round 5. In contrast, the baseline systems converge more slowly and finish at 3 failures each, indicating limited effectiveness in clarifying ambiguous task specifications.
Also, similar results can be observed in inter-agent misalignment (Fig. \ref{fig:3e}) and task verification (Fig. \ref{fig:3f}).

Compared to the baseline, SEMAP not only reduces failure counts more effectively, but also exhibits a more stable downward trend across rounds, whereas baseline failures often reappear after temporary drops.
The underlying mechanisms contributing to this stability are further discussed later.

\textit{Thus, RQ2 is answered: SEMAP promotes more reliable and consistent behavior across collaboration rounds, reducing the recurrence of specification, coordination, and verification failures in both function-level and deployment-level development.}

\section{Conclusion and Future Works}
\label{sec:conclusion}

This paper presents SEMAP, a protocol-layer methodology for addressing common failures in multi-agent LLMs, including under-specification, coordination misalignment, and inadequate verification.
SEMAP instantiates three SE-informed principles, behavioral contracts, structured messaging, and lifecycle-guided execution with verification, supporting both centralized and decentralized workflows.
Empirical results across diverse SE tasks demonstrate that SEMAP substantially improves system robustness.
In function-level and deployment-level code development, it achieves up to 69.6\% and 56.7\% reductions in total failures, with the greatest gains observed in mitigating under-specification and coordination issues.
In vulnerability detection, SEMAP consistently reduces failure rates across Python and C/C++ tasks up to 47.4\%, while promoting better agent alignment in decentralized workflows.

To strengthen validity, future experiments will be scaled to larger datasets, agent populations, and longer workflows, and compared against more baselines, including single-agent LLMs and domain-specific detectors.
Ablation studies will isolate the impact of contracts, messaging, and lifecycle control.
Future work also includes measuring resource overhead, enabling cross-agent tool use, adding formal protocol correctness verification, and releasing artifacts for reproducibility.

\bibliographystyle{IEEEtran}
\bibliography{APSEC2025}

\end{document}